\documentclass[pre,showpacs,showkeys,amsmath,amssymb,floatfix,eqsecnum]{revtex4}
\usepackage[T1]{fontenc}
\usepackage[latin9]{inputenc}
\usepackage{bm}
\usepackage{amsmath}
\usepackage{amssymb}
\usepackage{graphicx}
\usepackage{esint}

\makeatletter

\providecommand{\tabularnewline}{\\}

\@ifundefined{textcolor}{}
{%
 \definecolor{BLACK}{gray}{0}
 \definecolor{WHITE}{gray}{1}
 \definecolor{RED}{rgb}{1,0,0}
 \definecolor{GREEN}{rgb}{0,1,0}
 \definecolor{BLUE}{rgb}{0,0,1}
 \definecolor{CYAN}{cmyk}{1,0,0,0}
 \definecolor{MAGENTA}{cmyk}{0,1,0,0}
 \definecolor{YELLOW}{cmyk}{0,0,1,0}
}

\usepackage{bm}

\newcommand{\C}[1]{{}}

\makeatother

\begin{document}

\title{Thermodynamic Casimir Effect in the large-$n$ limit}

\author{Denis Comtesse}

\author{Alfred Hucht}

\author{Daniel Grüneberg}

\affiliation{Fachbereich Physik, Universität Duisburg-Essen, D-47048 Duisburg,
Germany}
\begin{abstract}
We consider systems with slab geometry of finite thickness $L$ that
undergo second order phase transitions in the bulk limit and belong
to the universality class of $O(n)$-symmetric systems with short-range
interactions. In these systems the critical fluctuations at the bulk
critical temperature $T_{\mathrm{c},\infty}$ induce a long-range
effective force called the ``thermodynamic Casimir force''. We describe
the systems in the framework of the $O(n)$-symmetric $\phi^{4}$-model,
restricting us to the large-$n$ limit $n\to\infty$. In this limit
the physically relevant case of three space dimensions $d=3$ can
be treated analytically in systems with translational symmetry as,
e.g., in the bulk or slabs with periodic or antiperiodic boundary
conditions. We consider Dirichlet and open boundary conditions at
the surfaces that break the translational invariance along the axis
perpendicular to the slab. From the broken translational invariance
we conclude the necessity to solve the systems numerically. We evaluate
the Casimir amplitudes for Dirichlet and open boundary conditions
on both surfaces and for Dirichlet on one and open on the other surface.
Belonging to the same surface universality class we find the expected
asymptotic equivalence of Dirichlet and open boundary conditions.
To test the quality of our method we confirm the analytical results
for periodic and antiperiodic boundary conditions. 
\end{abstract}



\maketitle

\section{Introduction \label{sec:intro}}

In a fluctuating medium confined between two surfaces long-range forces
may arise. This effect was first predicted by H.~B.~G.~Casimir
\cite{Cas48} who considered a confinement of the electromagnetic
vacuum fluctuations between two perfectly conducting metal plates.
The boundary conditions imposed by the two plates modify these fluctuations
in such a way that the energy of the system becomes dependent on the
separation of the plates. From this energy gradient in the separation
a long-range force, the so called Casimir force arises. This force
was experimentally measured for the first time in 1998 by Mohideen
and Roy \cite{MR98}. After this breakthrough more measurements followed
\cite{BCOR00,E00}, confirming Casimir's prediction with high accuracy.

An analogous effect is the so called \textit{thermodynamic} Casimir
effect which was first proposed by Fisher and de Gennes \cite{FdG78}.
It is caused by critical fluctuations of a medium near its bulk critical
point $T_{\mathrm{c},\infty}$, where these fluctuations are correlated
over an infinite range. The spatial confinement of such fluctuations
also gives rise to a long-range Casimir force.

The thermodynamic Casimir effect was experimentally proven for the
first time by Garcia and Chan in thin films of liquid $^{4}\mathrm{He}$
at its transition temperature to superfluidity \cite{GC99}. They
measured a characteristic dip of the film thickness at the critical
temperature, that can be attributed to the thermodynamic Casimir force.
The experimental data are excellently reproduced by Monte Carlo simulations
of the XY model with open boundary conditions \cite{Hucht07} which
is in the same universality class as the superfluid transition in
$^{4}\mathrm{He}$.

Due to the presence of strong fluctuations in critical systems there
are only a few exact results for the thermodynamic Casimir effect
yet. These are usually restricted to the case of periodic boundary
conditions along the slab. Though there are perturbation theory results
up to second order for experimentally relevant boundary conditions
as, e.g., Dirichlet boundary conditions \cite{KD92b}, exact results
for these are still lacking.

The remainder of this paper is organized as follows. In the next section
we carry the elementary features of the thermodynamic Casimir effect
together, and introduce the large-$n$ limit. In Section \ref{sec:fp}
the model considered in the present work will be defined and its excess
free energy will be derived in Section \ref{sec:FreeEnerg}. This
calculation leads to a self-consistent eigenvalue problem, which is
solved numerically, providing the possibility for numerical estimation
of the Casimir amplitudes. The paper is closed with a discussion of
the achieved results.

\subsection{Casimir effect}

As predicted by finite-size scaling theory a statistical system confined
between two parallel surfaces exhibits a dependence of the free energy
on the separation $L$ of the surfaces \cite{Bar83}. This $L$-dependence
leads to an effective force between the surfaces which is a typical
example of a finite-size effect. This thermodynamic Casimir force
is defined as \cite{FdG78,Kre94,Kre99,BDT00} 
\begin{equation}
{\mathcal{F}}_{C}(T,L)=-k_{B}T\,\frac{\partial f_{\text{ex}}(T,L)}{\partial L}\;,\label{eq:FCdef}
\end{equation}
where $f_{\text{ex}}$ denotes the reduced excess free energy per
unit area, which is given by $f_{\text{ex}}=f_{L}-Lf_{\text{b}}$
with the total free energy per unit area $f_{L}$ and the reduced
bulk free energy density $f_{\text{b}}$. The limit $A\to\infty$
of the cross sectional area has been taken.

For temperatures away from the bulk critical point the bulk correlation
length $\xi_{\infty}(T)$ is much smaller than the macroscopic separation
$L$. In this regime the Casimir force decays exponentially for large
$L$.

As it is generally known, $\xi_{\infty}(T)$ diverges as $|T-T_{\mathrm{c},\infty}|^{-\nu}$
at the bulk critical temperature $T_{\mathrm{c},\infty}$, with critical
exponent $\nu$, so in the vicinity of $T_{\mathrm{c},\infty}$ the
bulk correlation length is of the same order as $L$. This implies
that the Casimir force $\mathcal{F}_{C}(T,L)$ extends to distances
much larger than the microscopic scale $a$ ($\simeq$ lattice constant).
In this regime the Casimir force decays algebraically for large $L$
\cite{FdG78}.

Examining the excess free energy more precisely one finds that it
can be decomposed into two contributions 
\begin{equation}
f_{\text{ex}}(T,L)=f_{s}(T)+f_{\text{res}}(T,L)\;.\label{eq:delfex}
\end{equation}
The first one is the $L$-independent nonlocal surface excess free
energy $f_{s}(T)\equiv f_{\text{ex}}(T,\infty)$ \cite{Die86a}. The
second one is the residual finite-size contribution $f_{\text{res}}(T,L)$
which contains the whole information about the $L$-dependence of
the free energy.

At the bulk critical point $T_{\mathrm{c},\infty}$ the finite-size
contribution has the characteristic algebraic behavior %
\footnote{$\sim$ means ''asymptotically equal'', i.e., $f(x)\sim g(x)\Leftrightarrow\mathrm{lim}_{x\to\infty}f(x)/g(x)=1$%
} 
\begin{equation}
f_{\text{res}}(T_{\mathrm{c},\infty},L)\mathop{\sim}\limits _{L\to\infty}\Delta_{C}\, L^{-(d-1)},\label{eq:cadef}
\end{equation}
leading to a long-range effective Casimir force 
\begin{equation}
\frac{\mathcal{F}_{C}(T_{\mathrm{c},\infty},L)}{k_{B}T_{\mathrm{c},\infty}}\mathop{\sim}\limits _{L\to\infty}(d-1)\,\Delta_{C}\, L^{-d}.\label{eq:FCTc}
\end{equation}
The quantity $\Delta_{C}$ is an universal finite-size quantity, called
the Casimir amplitude \cite{FdG78}. Universal means that it depends
only on the bulk universality class of the phase transition and the
surface universality class of the boundary planes \cite{Die86a}.
It is in particular independent of microscopic details of the system,
such as, e.g., the lattice structure.

Near the bulk critical point the residual free energy density and
the Casimir force obey the scaling forms 
\begin{equation}
f_{\text{res}}(T,L)\sim L^{-(d-1)}\,\Theta(L/\xi_{\infty}(T))\label{eq:scffintro}
\end{equation}
and 
\begin{equation}
\frac{\mathcal{F}_{C}(T,L)}{k_{B}T}\sim L^{-d}\,\vartheta(L/\xi_{\infty}(T)),\label{eq:scfFCintro}
\end{equation}
where $\Theta(x)$ and $\vartheta(x)$ are universal functions with
the scaling argument $x=L/\xi_{\infty}$. At $T_{\mathrm{c},\infty}$,
where $L/\xi_{\infty}=0$, they take the values $\Theta(0)=\Delta_{C}$
and $\vartheta(0)=(d-1)\Delta_{C}$.

There is no complete analytic theory for the thermodynamic Casimir
effect joining the regimes above and below $T_{\mathrm{c},\infty}$.
In the framework of the $\phi^{4}$-theory perturbative calculations
up to second order for different boundary conditions at and above
$T_{\mathrm{c},\infty}$ \cite{KD92b,DGS06,GD08,Fe08} were performed.
There are also exact results for the spherical model with periodic
\cite{BDT00} and antiperiodic \cite{Dan09} boundary conditions.
However, for slab systems with non-periodic boundary conditions no
exact results are available. This is the aim of the present work.

\subsection{The large-$n$ limit and the spherical model}

To investigate the thermodynamic Casimir effect one needs a model
that shows the characteristic critical properties of a second order
phase transition. There are various classes of models that exhibit
critical behavior but here we restrict ourselves to the class of classical
$O(n)$-invariant $n$-vector lattice models. Those can be represented
by the Hamiltonian 
\begin{equation}
\mathcal{H}_{n}=-\frac{J}{2}\sum_{\langle i,j\rangle}\boldsymbol{S}_{i}\cdot\boldsymbol{S}_{j},
\end{equation}
where the $\boldsymbol{S}=(s_{1},...,s_{n})$ are $n$-component vector
spin variables, restricted to the local constraint $|\boldsymbol{S}|^{2}=1$.
The exchange coupling $J$ is assumed to have a positive nonzero value
and the sum runs over all nearest neighbors.

In the limit $n\to\infty$, which is called the large-$n$ limit this
model can be solved analytically for the bulk or for systems of finite
size \cite{BZ82}. This case is connected to the so called the spherical
model, which was first introduced and solved by Berlin and Kac \cite{BerKac52}
and is defined by the Hamiltonian 
\begin{equation}
\mathcal{H}_{\mathrm{SM}}=-\frac{J'}{2}\sum_{\langle i,j\rangle}s_{i}s_{j}.
\end{equation}
Here the $s_{i}$ are scalar spin variables fulfilling a global constraint
$\sum_{i}s_{i}^{2}=N$ called ``spherical constraint'', where $N$
is the number of lattice sites.

This model is even more easy to handle if one uses the constraint
$\left<\sum_{i}s_{i}^{2}\right>=N$ where one takes the statistical
mean over the sum. It is then called the ``mean spherical model''
but the both formulations are equivalent in translationally invariant
systems.

It was found by Stanley \cite{Sta68} that these spherical models
are equivalent to the large-$n$ limit for translationally invariant
systems as, e.g., the bulk case or systems with periodic boundary
conditions.

To hold this equivalence in non-translationally invariant systems
the spherical constraint has to be modified \cite{Knops73}. In the
case of a slab where the boundary conditions are non periodic one
needs a spherical constraint for each layer parallel to the surfaces.
These multiple constraints make the analytical treatment impossible
so that numerical calculations are needed. This is the challenge we
are going to face here.

The large-$n$ limit can also be understand as the zeroth order of
a systematical expansion in $1/n$ \cite{Ma76}. This connects our
theory to the physical relevant case of, e.g., three order parameter
components known as the Heisenberg model. But this expansion is more
useful for formal studies than to comparison to experimental observations.

\section{Definition of the model \label{sec:fp}}

In the last chapter we have discussed $O(n)$-symmetrical lattice
models but now we turn to the description of the universal critical
behavior in the framework of an $O(n)$-symmetrical $\phi^{4}$-model.
The connection between both is that the $n$-vector lattice model
can be mapped on the $n$-vector $\phi^{4}$-model by means of coarse
graining \cite{CL95}.

We consider a $d$-dimensional slab of finite thickness $L$, which
occupies the volume $\mathfrak{V}=\mathbb{R}^{d-1}\times[0,L]$. Let
the $x_{j}$ with $\; j=1\dotsc,d,$ be cartesian coordinates and
we write $x_{d}\equiv z$ to denote the coordinate of the finite direction
called the perpendicular direction. The full position vector is given
by $\bm{x}=(\bm{y},z)$, where $\bm{y}=(x_{1},\dotsc,x_{d-1})$ denotes
the coordinates of the directions along the slab called parallel directions.

As discussed in detail in \cite{Die86a} the general Hamiltonian of
$\phi^{4}$-models with surfaces is given by 
\begin{equation}
\mathcal{H}[\bm{\phi}(\boldsymbol{x})]=\int_{\mathfrak{V}}\mathcal{L}_{\mathfrak{V}}[\bm{\phi}(\boldsymbol{x})]\,\mathrm{d}V+\int_{\partial\mathfrak{V}}\mathcal{L}_{\partial\mathfrak{V}}[\bm{\phi}(\boldsymbol{x})]\,\mathrm{d}A.\label{eq:Hform}
\end{equation}

The densities $\mathcal{L}_{\mathfrak{V}}(\bm{x})$ and $\mathcal{L}_{\partial\mathfrak{V}}(\bm{x})$
depend on the $n$-component order parameter field $\bm{\phi}(\bm{x})$
and its derivatives. This order parameter field is a classical $n$-component
vector field that, using the language of magnetism, can be interpreted
as the local spin density. The first part is the bulk Hamiltonian
and the second is the additional surface Hamiltonian. The latter is
taken in account because the coupling at the surface might deviate
from that in the bulk.

The bulk density $\mathcal{L}_{\mathfrak{V}}(\bm{x})$ is given by
\begin{equation}
\mathcal{L}_{\mathfrak{V}}[\bm{\phi}(\bm{x})]=\frac{1}{2}\,\sum_{\alpha=1}^{n}(\nabla\phi_{\alpha}(\bm{x}))^{2}+\frac{\mathring{\tau}}{2}\,\boldsymbol{\phi}^{2}(\bm{x})+\frac{\mathring{u}}{4!n}\,\left(\boldsymbol{\phi}^{2}(\bm{x})\right)^{2},\label{eq:Lb}
\end{equation}
where $\mathring{\tau}$ is the temperature variable often called
mass and $\mathring{u}$ is the coupling constant.

Note that in contrast to $\phi^{4}$-theories with a finite number
of order parameter field components there is an additional $1/n$
in front of the $\phi^{4}$-term, which is needed to make the limit
$n\to\infty$ well defined \cite{Ami05}.

The general boundary density $\mathcal{L}_{\partial\mathfrak{V}}(\bm{x})$
reads 
\begin{equation}
\mathcal{L}_{\partial\mathfrak{V}}[\bm{\phi}(\boldsymbol{x})]=\frac{\mathring{c}(\bm{x})}{2}\,\boldsymbol{\phi}^{2}(\bm{x}).\label{eq:L1}
\end{equation}
The $\mathring{c}(\bm{x})$ is the surface enhancement that controls
the additional couplings at the surface. It may have different values
on the two boundaries $\partial\mathfrak{B}_{1}$ and $\partial\mathfrak{B}_{2}$,
i.e., 
\begin{equation}
\mathring{c}(\bm{x})=\begin{cases}
\mathring{c}_{1} & \text{for }\bm{x}\in\partial\mathfrak{B}_{1},\\
\mathring{c}_{2} & \text{for }\bm{x}\in\partial\mathfrak{B}_{2}.
\end{cases}\label{eq:c12}
\end{equation}

In our framework these boundary terms need not to be taken into account
any longer because they give no contribution to the finite size energy
that we are interested in and one can take account for the boundary
conditions on the surface without those additional terms. The boundary
conditions will be taken into account in the calculation of the eigensystem
of the system. How this is done will be described in the next section.

\section{The free energy}

\label{sec:FreeEnerg}

We now want to derive the exact expression for the free energy of
the system. In the large-$n$ limit such an expression can be obtained
in different ways. One can use, e.g., variational techniques as well
as the saddle-point approximation \cite{Moshe}. Both become exact
in the limit $n\to\infty$. In the following we resort on the saddle-point
approximation method.

We start the calculation by computing the canonical partition function
$\mathcal{Z}$ of the system. It is given by the functional integral
over all possible configurations of the order parameter field $\boldsymbol{\phi}(\boldsymbol{x})$.
We write 
\begin{equation}
\mathcal{Z}=\int\mathcal{D}[\boldsymbol{\phi}]e^{-\mathcal{H}[\boldsymbol{\phi}]}=\int\mathcal{D}[\boldsymbol{\phi}]\exp\left\{ -\int\mathrm{d}^{d}\boldsymbol{x}\left[\frac{1}{2}\,\sum_{\alpha=1}^{n}(\nabla\phi_{\alpha})^{2}+\frac{\mathring{\tau}}{2}\,\boldsymbol{\phi}^{2}+\frac{\mathring{u}}{4!n}\,\left(\boldsymbol{\phi}^{2}\right)^{2}\right]\right\} .
\end{equation}
To determine the partition function with the saddle-point approximation
we transform the last expression by applying the Hubbard-Stratonovich
decoupling technique \cite{Huba59} and introducing an additional
parameter $\psi$. This leads to 
\begin{equation}
\mathcal{Z}=C\int\mathcal{D}[\boldsymbol{\phi}]\int\mathcal{D}[i\psi]\exp\left\{ -\frac{1}{2}\int\mathrm{d}^{d}\boldsymbol{x}\left[\boldsymbol{\phi}(\boldsymbol{x})(-\nabla^{2}+\mathring{\tau}-\psi(z))\boldsymbol{\phi}(\boldsymbol{x})-\frac{3n}{\mathring{u}}\psi^{2}(z)\right]\right\} .
\end{equation}
The constant pre-factor $C$ is irrelevant since it does not contribute
to the excess free energy. The introduced variational parameter $\psi$
is chosen to depend on $z$ because the boundary conditions break
the translational invariance along that axis.

The integration over $\boldsymbol{\phi}(\boldsymbol{x})$ is now Gaussian
and can be carried out (see e.g. \cite{GOLD92}). This leads us to
the expression 
\begin{equation}
\mathcal{Z}=\tilde{C}\int\mathcal{D}[i\psi]\exp\left[-\frac{n}{2}\mathrm{Tr}\log(-\nabla^{2}+\mathring{\tau}-\psi(z))+\frac{3n}{2\mathring{u}}\int_{0}^{L}\mathrm{d}z\,\psi^{2}(z)\right].
\end{equation}

The remaining functional integral can be evaluated by the saddle-point
approximation. This means that we take the value of the integrand
at its saddle point as solution of the integral. The number of order
parameter components $n$ serves as large parameter in the exponential
function that makes the saddle-point a sharp peak as required for
this approximation. With growing $n$ the peak becomes sharper so
that the approximation gets better and will become exact in the limit
$n\to\infty$. So we conclude that the partition function is given
by 
\begin{equation}
\mathcal{Z}=\bar{C}\exp\left[-\frac{n}{2}\mathrm{Tr}\log(-\nabla^{2}+\mathring{\tau}-\psi_{0}(z))+\frac{3n}{2\mathring{u}}\int_{0}^{L}\mathrm{d}z\,\psi_{0}^{2}(z)\right].\label{eq:PartFkt1}
\end{equation}

From the partition function we obtain the reduced free energy using
its definition $F(L)=\mathcal{F}/k_{\mathrm{B}}T=-\log\mathcal{Z}$.
It is useful to consider the reduced free energy density per cross
sectional area and order parameter component, $f_{L}=F/An$ with $A\to\infty$,
that comes out as 
\begin{equation}
f_{L}=f_{0}+\frac{1}{2}\int_{\boldsymbol{p}}^{(d-1)}\int_{0}^{L}\mathrm{d}z\,\langle z|\log(-\partial_{z}^{2}+\boldsymbol{p}^{2}+\mathring{\tau}-\psi_{0}(z))|z\rangle-\frac{3}{2\mathring{u}}\int_{0}^{L}\mathrm{d}z\,\psi_{0}^{2}(z),\label{eq:PartFkt2}
\end{equation}
where we used the Dirac notation $\mathrm{Tr}(\circ)=\int_{0}^{L}\mathrm{d}z\langle z|\circ|z\rangle$
and defined 
\begin{equation}
\int_{\boldsymbol{p}}^{(d-1)}\equiv\prod_{i=1}^{d-1}\int_{-\infty}^{\infty}\frac{\mathrm{d}p_{i}}{2\pi}.
\end{equation}
Note that we have set $\psi=\psi_{0}$ in equations (\ref{eq:PartFkt1})
and (\ref{eq:PartFkt2}) where $\psi_{0}$ is the saddle-point value.
It is defined by requiring that at the saddle-point the integrand
must be extremal in $\psi$, and obeys the self-consistent equation
\begin{equation}
\psi_{0}(z)=-\frac{\mathring{u}}{6}\int_{\boldsymbol{p}}^{(d-1)}\langle z|\frac{1}{-\partial_{z}^{2}+\boldsymbol{p}^{2}+\mathring{\tau}-\psi_{0}(z)}|z\rangle.\label{eq:saddelpnt}
\end{equation}
We now examine this self-consistent equation and determine the saddle-point
value and the eigensystem along the $z$ axis. But first we want to
remark the following: The problem can also be approached with the
help of diagrammatic perturbation theory. From this theory one obtains
the Dyson equation 
\begin{equation}
\left[\tilde{G}(\boldsymbol{k})\right]^{-1}=\boldsymbol{k}^{2}+\mathring{\tau}-\tilde{\Sigma}(\boldsymbol{k})
\end{equation}
for the two point correlation function of the system often named propagator
$\tilde{G}(\boldsymbol{k})$. The contributions from all orders of
perturbation theory are summed in the mass-operator $\Sigma(\boldsymbol{k})$.
Usually this cannot be done without approximation but, in the large-$n$
limit the contributions to the mass operator can be summed up exactly
(see \cite{Ami05,BM77}). That mass-operator has the same self-consistent
structure as $\psi_{0}(z)$ in equation (\ref{eq:saddelpnt}). So
in comparison we can identify $\psi_{0}(z)=\Sigma(z)$ and we will
use $\Sigma(z)$ instead of $\psi_{0}(z)$ the following. For simplicity
we write the mass-operator $\Sigma(z)$ in the mixed $\boldsymbol{p}$-$z$-representation
\begin{equation}
\Sigma(z)=-\frac{\mathring{u}}{6}\int_{\boldsymbol{p}}^{(d-1)}\sum_{\nu}\frac{\hat{\varphi}_{\nu}(z)\hat{\varphi}_{\nu}^{*}(z)}{\hat{\epsilon}_{\nu}+\boldsymbol{p}^{2}+\mathring{\tau}}.
\end{equation}
The $\hat{\varphi}_{\nu}(z)$ and the $\hat{\epsilon}_{\nu}$ are
the eigenfunctions and the eigenvalues used to expand the $z$-dependence
and are given by the equation 
\begin{equation}
\left[-\partial_{z}^{2}-\Sigma(z)\right]\hat{\varphi}_{\nu}(z)=\hat{\epsilon}_{\nu}\hat{\varphi}_{\nu}(z).
\end{equation}

Obviously the quantity $\Sigma(z)$ is not well defined because the
$\boldsymbol{p}$-integration is UV-divergent. Usually one would have
to deal with the procedures of regularization and renormalization
to give the divergent expression a physical meaning. But our case
is much simpler because we are not interested in the mass-operator
itself. Thus we define another quantity which is UV-convergent by
taking the difference of $\Sigma(z)$ and a quantity which is UV-divergent
to the same order.

For this purpose we decompose the mass $\mathring{\tau}$ into $\mathring{\tau}=\mathring{\tau}_{\mathrm{c}}+\delta\mathring{\tau}$.
The $\mathring{\tau}_{\mathrm{c}}$ gives the value of $\mathring{\tau}$
at which the corresponding bulk system reaches criticality and $\delta\mathring{\tau}$
gives the deviation from this critical value. The $\mathring{\tau}_{\mathrm{c}}$
is given by the expression \cite{Ami05} 
\begin{equation}
\mathring{\tau}_{\mathrm{c}}=-\frac{\mathring{u}}{6}\int_{\boldsymbol{p}}^{(d-1)}\int_{-\infty}^{\infty}\frac{\mathrm{d}q}{2\pi}\frac{1}{\boldsymbol{p}^{2}+q^{2}}.
\end{equation}
It is UV-divergent to the same order as $\Sigma(z)$. From the sum
of both we now obtain a UV-convergent expression. We define the new
quantity 
\begin{equation}
U(z)=-\Sigma(z)+\mathring{\tau}_{\mathrm{c}}=\frac{\mathring{u}}{6}\int_{\boldsymbol{p}}^{(d-1)}\left(\sum_{\nu}\frac{\varphi_{\nu}(z)\varphi_{\nu}^{*}(z)}{\epsilon_{\nu}+\boldsymbol{p}^{2}+\delta\mathring{\tau}}-\int_{-\infty}^{\infty}\frac{\mathrm{d}q}{2\pi}\frac{1}{\boldsymbol{p}^{2}+q^{2}}\right).\label{eq:UvonZ}
\end{equation}
Introducing this quantity the eigensystem is given by 
\begin{equation}
\left[-\partial_{z}^{2}+U(z)\right]\varphi_{\nu}(z)=\epsilon_{\nu}\varphi_{\nu}(z).\label{eq:EW_Gl_U}
\end{equation}

To simplify the further discussion we will interpret the two terms
on the right hand side of equation (\ref{eq:UvonZ}) analogously to
Ref.~\cite{BM77}. The first term on the right hand side is the propagator
of the slab system for temperature deviations from the critical point
defined by $\delta\mathring{\tau}$. The second term is the propagator
of the corresponding bulk system at criticality where $\delta\mathring{\tau}=0$.
With this definitions one may write equation (\ref{eq:UvonZ}) in
the following form 
\begin{equation}
U(z)=\frac{\mathring{u}}{6}\int_{\boldsymbol{p}}^{(d-1)}\left(\tilde{G}(\boldsymbol{p};z,z)-\int_{-\infty}^{\infty}\frac{\mathrm{d}q}{2\pi}\tilde{G}_{\mathrm{bulk}}(\boldsymbol{p};q)\right).
\end{equation}

Equations (\ref{eq:UvonZ}) and (\ref{eq:EW_Gl_U}) form a self-consistent
eigenvalue problem similar to the one dimensional Schrödinger equation
of a particle in a potential $U(z)$. The numerical solution of this
problem is the content of the following subsection.

\subsection{Discretization of the problem}

Using a numerical method we now calculate the potential $U(z)$ and
the corresponding eigensystem self-consistently. Therefore the model
needs to be discrete along the $z$-axis. We introduce a discretization
of the $z$-direction with $L$ points and set the lattice constant
$a=1$ without loss of generality, such that $z\to z_{i}$ with $z_{i}\in\{1,2,3,...,L\}$. 

This discretization leads to some more changes. At first the number
of eigenvalues $\epsilon_{\nu}$ and functions $\varphi_{\nu}$ is
restricted to the number of discrete points $L$. At second there
is also a change of the dispersion relation in the corresponding bulk
model. The $q^{2}$ in the second term of the right hand side of equation
(\ref{eq:UvonZ}) has to be replaced by the dispersion relation 
\begin{equation}
\epsilon(q)=4\sin^{2}\left(\frac{q}{2}\right)\label{eq:DispRel}
\end{equation}
of a lattice theory. Due to the periodicity of (\ref{eq:DispRel})
the integration with respect to $q$ is restricted to the first Brillouin
zone. Implementing these modifications, the potential $U(z)$ displayed
in (\ref{eq:UvonZ}) becomes 
\begin{equation}
U(z_{i})=\frac{\mathring{u}}{6}\int_{\boldsymbol{p}}^{(d-1)}\left[\sum_{\nu=1}^{L}\frac{\varphi_{\nu}(z_{i})\varphi_{\nu}^{*}(z_{i})}{\epsilon_{\nu}+\boldsymbol{p}^{2}+\delta\mathring{\tau}}-\int_{0}^{2\pi}\frac{\mathrm{d}q}{2\pi}\frac{1}{\boldsymbol{p}^{2}+4\sin^{2}(q/2)}\right],
\end{equation}
where the corresponding eigenvalue equation (\ref{eq:EW_Gl_U}) now
reads 
\begin{equation}
\left[-\boldsymbol{\partial}_{z}^{2}+\boldsymbol{U}\right]\boldsymbol{\varphi}_{\nu}=\epsilon_{\nu}\boldsymbol{\varphi}_{\nu}.
\end{equation}
Here the $\boldsymbol{\partial}_{z}^{2}$ and $\boldsymbol{U}$ are
finite $L\times L$ matrices and the $\boldsymbol{\varphi}_{\nu}$
are the corresponding finite $L$-component vectors. We specify these
objects now: The $\boldsymbol{\partial}_{z}^{2}$ stands for the symmetric
differentiation operator for the second derivative of a discrete function.
Usually this operator is defined by 
\begin{equation}
\boldsymbol{\partial}_{z}^{2}\varphi_{\nu}(z_{i})=\varphi_{\nu}(z_{i+1})-2\varphi_{\nu}(z_{i})+\varphi_{\nu}(z_{i-1})
\end{equation}
 and can be represented by the matrix 
\begin{equation}
\boldsymbol{\partial}_{z}^{2}=\left(\begin{array}{cccccc}
-x & 1 & 0 & \dots & 0 & y\\
1 & -2 & 1 &  &  & 0\\
0 & 1 & -2 & \ddots &  & \vdots\\
\vdots &  & \ddots & \ddots &  & \vdots\\
0 &  &  &  & -2 & 1\\
y & 0 & \dots & \dots & 1 & -x
\end{array}\right).
\end{equation}

At this stage the boundary conditions of the system come into play.
We can control them by choosing adequate values for $x$ and $y$.
The reader can convince himself that if one chooses $x=1$ {[}$x=2${]}
and $y=0$ this corresponds to open {[}Dirichlet{]} boundary conditions.
If one sets $y=1$ {[}$y=-1${]} and $x=2$ one gets periodic {[}antiperiodic{]}
boundary conditions.

The matrix $\boldsymbol{U}$ is the diagonal matrix of the $U(z_{i})$
\begin{equation}
\boldsymbol{U}=\mathrm{diag}\left[U(z_{i})\right]
\end{equation}
and the eigenfunctions of the discrete model are represented as vectors
where the components are given by 
\begin{equation}
\varphi_{\nu}(z_{i})=\varphi_{\nu,i}.
\end{equation}

In this discrete form the self-consistent eigenvalue problem can be
implemented into an iterative routine to determine the eigensystem
and the matrix $\boldsymbol{U}$. The calculations were carried out
with build-in functions of the program \textsc{Mathematica} \cite{MATMA}.

\begin{figure}
\begin{centering}
\includegraphics[clip,width=0.49\columnwidth]{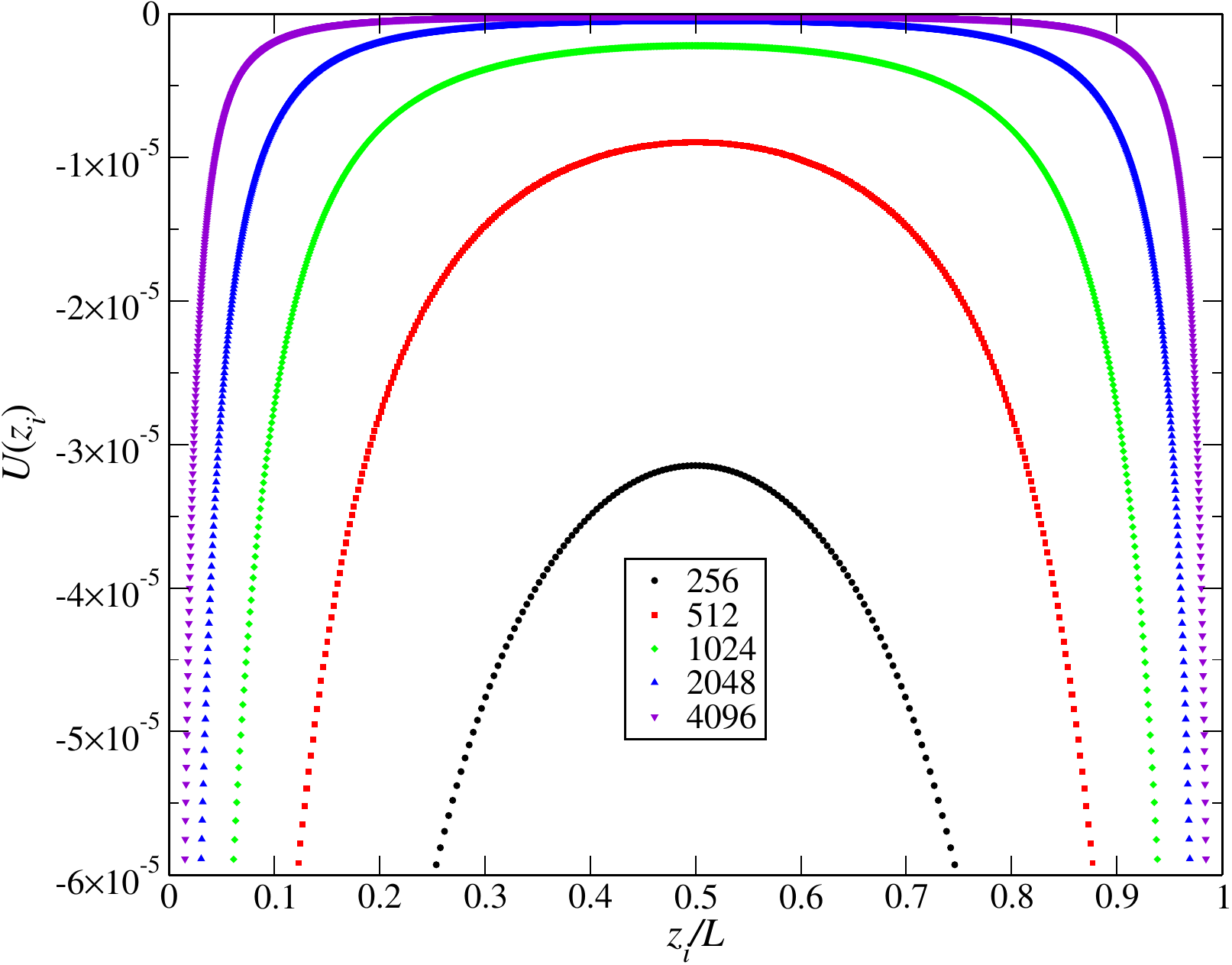}\hfill{}\includegraphics[clip,width=0.49\columnwidth]{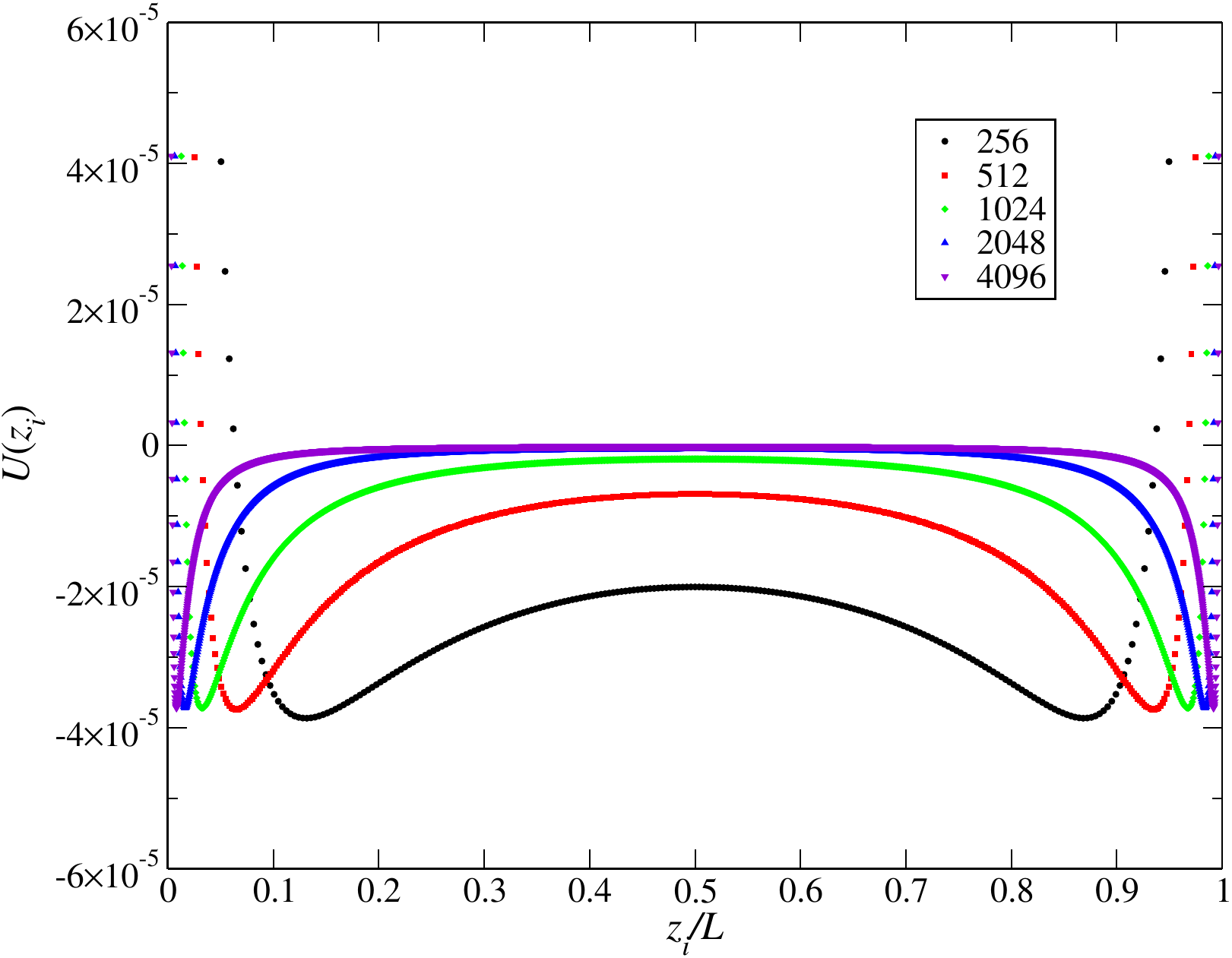}
\par\end{centering}

~

\begin{centering}
\includegraphics[clip,width=0.49\columnwidth]{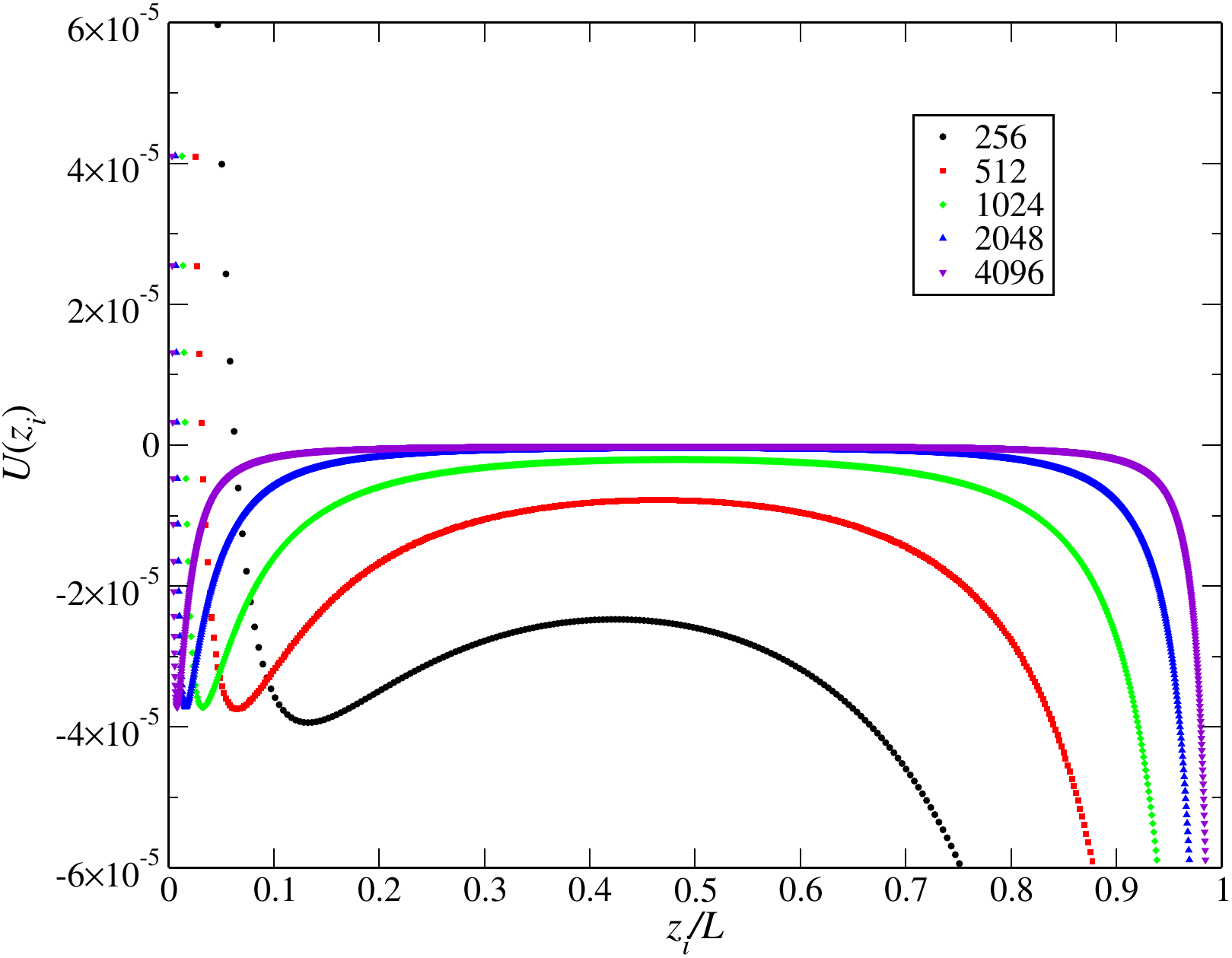}
\par\end{centering}

\centering{}\caption{Numerical data of the potential $U(z_{i})$ for systems with Dirichlet-Dirichlet
(left), open-open (right), and open-Dirichlet (bottom) boundary conditions.
\label{fig:DD_DO_OO} }
\end{figure}
The iteration starts from an initial guess for the potential. From
this guess the associated eigensystem is computed. With the eigensystem
one computes again the potential for the next iteration step. This
procedure is repeated until the difference between the steps is small
enough to fulfill a certain termination condition.

Figure \ref{fig:DD_DO_OO} shows the numerical results for the potential
$U(z_{i})$ plotted over the scaling variable $z_{i}/L$. We show
system sizes $L$ in powers $2^{n}$ from $L=256$ to $L=4096$, and
we set $\mathring{u}=1$ in to following. Figure \ref{fig:DD_DO_OO}(left)
shows the results for the systems with Dirichlet boundary conditions
on both surfaces. The potential as anticipated diverges at the boundaries
to ensure that the eigenfunctions fall to zero on the surface. In
the middle of the slab the potential tends with a characteristic algebraic
behavior to its bulk value which is zero at criticality (see \cite{Ma76}).
This tendency is more and more distinct for bigger systems because
the influence of the boundaries in the middle of the slab decays with
increasing system size. Figure \ref{fig:DD_DO_OO}(right) shows the
results for open boundary conditions. In the middle of the slab the
potential approaches the behavior of systems with Dirichlet boundary
conditions. This approach becomes stronger for larger systems. But
next to the surfaces the evolution differs very strongly from the
Dirichlet case. The most remarkable difference is that the potential
has now a finite, positive value at the surface. When there are both
boundary conditions mixed in one system as it is shown in Figure \ref{fig:DD_DO_OO}(bottom)
we find the diverging behavior on the Dirichlet side and the finite
values on the open side. The values of the potential in the middle
of the slab are still converging to the bulk value with increasing
system size. 

The Dirichlet and the open system are that similar because they are
in the same surface universality class \cite{Die86a}. This means
that their asymptotical behavior for large $L$ and large values of
$z$ far away from the surface is identical. 

To test the quality of our method we have also computed the potential
$U(z)$ for the known cases of periodic and antiperiodic boundary
conditions with our numerical method. We found that our results are
in excellent agreement with the analytical results. A comparison between
our results for the Casimir amplitude calculated from this potentials
and the analytic results is shown in the next section.

\section{Excess free energy and Casimir amplitudes}

Now that the $U(z)$ is determined we can calculate the Casimir amplitudes.
But first we have to modify the expression of the free energy by using
the definitions $\mathring{\tau}=\mathring{\tau}_{\mathrm{c}}+\delta\mathring{\tau}$
and $\Sigma(z)=\mathring{\tau}_{\mathrm{c}}-U(z)$ to have only well
defined quantities in it.

As we are not interested in the free energy itself but in the difference
of the free energy of the slab and the bulk, called excess free energy
we need an expression for the free energy of the bulk. We obtain this
expression from a calculation analogous to that in the last chapter.
The translational invariance of the bulk allows one to obtain its
free energy analytically \cite{Ma76}.

Now the exact expression for the excess free energy defined in Sec.~\ref{sec:intro}
can be written in the form 
\begin{eqnarray}
f_{\mathrm{ex}} & = & \frac{1}{2}\int_{\boldsymbol{p}}^{(d-1)}\sum_{\nu}\log(\epsilon_{\nu}+\boldsymbol{p}^{2}+\delta\mathring{\tau})-\frac{3}{2\mathring{u}}\int_{0}^{L}\mathrm{d}z(\mathring{\tau}_{\mathrm{c}}-U(z))^{2}\nonumber \\
 &  & {}-\frac{L}{2}\int_{\boldsymbol{p}}^{(d-1)}\int_{-\infty}^{\infty}\frac{\mathrm{d}q}{2\pi}\log(\boldsymbol{p}^{2}+q^{2}+\delta\mathring{\tau}+U_{\mathrm{bulk}})+\frac{3L}{2\mathring{u}}(\mathring{\tau}_{\mathrm{c}}-U_{\mathrm{bulk}})^{2}.
\end{eqnarray}

Because of the discrete numerical evaluation of the potential $U(z)$
all $z$ dependencies in the last expression have to be discrete.
This is what will be done the next section.

\subsection{Discrete form of $\boldsymbol{f_{\mathrm{ex}}}$}

We now obtain the form of the excess free energy with a discretized
$z$ dependence. This means substituting $z\to z_{i}$ like in Sec.~\ref{sec:FreeEnerg}.
With this change we have to replace the dispersion relation of the
continuum by that of the lattice and restrict the $q$ momentum integrations
to the first Brillouin zone (see also Sec.~\ref{sec:FreeEnerg}).
Finally the real space integration along the $z$-axis has to be turned
into a sum. This is achieved using $\int_{0}^{L}\mathrm{d}z\to\sum_{i=1}^{L}$
as we have set $a=1$. From this changes we obtain the excess free
energy in the discrete form given by 
\begin{eqnarray}
f_{\mathrm{ex}} & = & \frac{1}{2}\int_{\boldsymbol{p}}^{(d-1)}\sum_{\nu=1}^{L}\log\left(\epsilon_{\nu}+\boldsymbol{p}^{2}+\delta\mathring{\tau}\right)-\frac{3}{2\mathring{u}}\sum_{i=1}^{L}(\mathring{\tau}_{\mathrm{c}}-U(z_{i}))^{2}\nonumber \\
 &  & {}-\frac{L}{2}\int_{\boldsymbol{p}}^{(d-1)}\int_{0}^{2\pi}\log\left[\boldsymbol{p}^{2}+4\sin^{2}\left(\frac{q}{2}\right)+\delta\mathring{\tau}+U_{\mathrm{bulk}}\right]+\frac{3L}{2\mathring{u}}\left(\mathring{\tau}_{\mathrm{c}}-U_{\mathrm{bulk}}\right)^{2}.
\end{eqnarray}

\begin{table}[b]
\begin{centering}
\begin{tabular}{|c||r@{\extracolsep{0pt}.}l|r@{\extracolsep{0pt}.}l|r@{\extracolsep{0pt}.}l|r@{\extracolsep{0pt}.}l|}
\hline 
BC & \multicolumn{2}{c|}{$a_{0}$} & \multicolumn{2}{c|}{$a_{2}$} & \multicolumn{2}{c|}{$a_{3}$} & \multicolumn{2}{c|}{$a_{4}$}\tabularnewline
\hline 
\hline 
OO & 0&0297 & $-$0&0122 & 0&0106 & $-$6&06~~~\tabularnewline
\hline 
DO & $-$0&0466 & $-$0&0119 & $-$2&97 & 441&\tabularnewline
\hline 
DD & ~~~~~0&0309~~~~ & $-$0&0115~~~~ & ~~~~~$-$4&53~~~~~ & ~~~~~427&\tabularnewline
\hline 
PBC & \multicolumn{2}{c|}{--} & $-$0&1530 & $-$1&23 & 51&0\tabularnewline
\hline 
~~APBC~~ & \multicolumn{2}{c|}{--} & ~~~~~0&2742 & $-$27&4 & 3390&\tabularnewline
\hline 
\end{tabular}
\par\end{centering}

\caption{Fit parameters for the excess free energy, Eq.~(\ref{eq:fit}), for
$\mathring{u}=1$. OO means open boundary conditions on both sides,
DD means the same with Dirichlet boundary conditions, and DO means
the mix of both. \label{tab:FitTab} }
\end{table}
Since we are interested in the Casimir amplitude we limit the following
discussion to the case $T=T_{\mathrm{c},\infty}$ and set the temperature
variable $\delta\mathring{\tau}=0$. In the following we use the result
that the $U_{\mathrm{bulk}}$ equals zero at $T_{\mathrm{c},\infty}$
(see \cite{Ma76}).

After evaluating the momentum integrals we get the expression \C{
\begin{equation}
f_{\mathrm{ex}}=\frac{2^{-d}\pi^{\frac{3-d}{2}}}{\Gamma\!\left(\frac{d+1}{2}\right)\cos\left(\frac{\pi d}{2}\right)}\sum_{\nu=1}^{L}\left[-\epsilon_{\nu}^{\frac{d-1}{2}}+\frac{2^{d-1}\Gamma\!\left(\frac{d}{2}\right)}{\sqrt{\pi}\Gamma\!\left(\frac{d+1}{2}\right)}\right]+\frac{\pi^{1-\frac{d}{2}}\Gamma\!\left({\textstyle \frac{d}{2}}-1\right)}{8\Gamma\!\left(\frac{d-1}{2}\right)^{2}\cos\left(\frac{\pi d}{2}\right)}\sum_{i=1}^{L}U(z_{i})-\frac{3}{2\mathring{u}}\sum_{i=1}^{L}U(z_{i})^{2}.
\end{equation}
ODER }
\begin{equation}
f_{\mathrm{ex}}=\frac{\pi^{1-\frac{d}{2}}}{\cos\left(\frac{\pi d}{2}\right)}\left[-\frac{2^{-d}\sqrt{\pi}}{\Gamma\!\left(\frac{d+1}{2}\right)}\sum_{\nu=1}^{L}\epsilon_{\nu}^{\frac{d-1}{2}}+\frac{L\Gamma\!\left(\frac{d}{2}\right)}{2\Gamma\!\left(\frac{d+1}{2}\right)^{2}}+\frac{\Gamma\!\left({\textstyle \frac{d}{2}}-1\right)}{8\Gamma\!\left(\frac{d-1}{2}\right)^{2}}\sum_{i=1}^{L}U(z_{i})\right]-\frac{3}{2\mathring{u}}\sum_{i=1}^{L}U(z_{i})^{2}.
\end{equation}
Examining this expression one recognizes that it has a remaining pole
term in $d=3$ space dimensions. This pole term can be understood
by realizing that the $q$-integration is carried out but the corresponding
summation over the eigenvalues is left unevaluated. The pole term
needs to be removed to set $d=3$ because the evaluation of the remaining
sum in arbitrary dimensions is not feasible.

To make the pole-term visible and isolate it we expand $f_{\mathrm{ex}}$
around space dimension $d=3$. In this expansion we find a term of
order $\mathcal{O}(1/(d-3))$ and a term of order $\mathcal{O}(1)$,
\begin{equation}
\frac{1}{8\pi}\left(\frac{2}{d-3}+\gamma-\log(4\pi)\right)\left\{ -\sum_{\nu=1}^{L}\epsilon_{\nu}+\sum_{i=1}^{L}[2+U(z_{i})]\right\} ,
\end{equation}
which vanish as both sums in the curly braces are equal to the trace
of the matrix $-\boldsymbol{\partial}_{z}^{2}+\boldsymbol{U}$. The
zeroth order of the expansion contains the whole information of the
excess free energy at $d=3$ so we restrict us to the discussion of
this term. The final result at $d=3$ becomes 
\begin{equation}
f_{\mathrm{ex}}=\frac{1}{8\pi}\sum_{\nu=1}^{L}\epsilon_{\nu}\big(1-\log\epsilon_{\nu}\big)-\frac{3}{2\mathring{u}}\sum_{i=1}^{L}U(z_{i})^{2}.
\end{equation}
 In this expression the remaining sums can be evaluated to obtain
the excess free energies for the given system sizes.

\begin{figure}
\centering{}\includegraphics[clip,width=0.48\columnwidth]{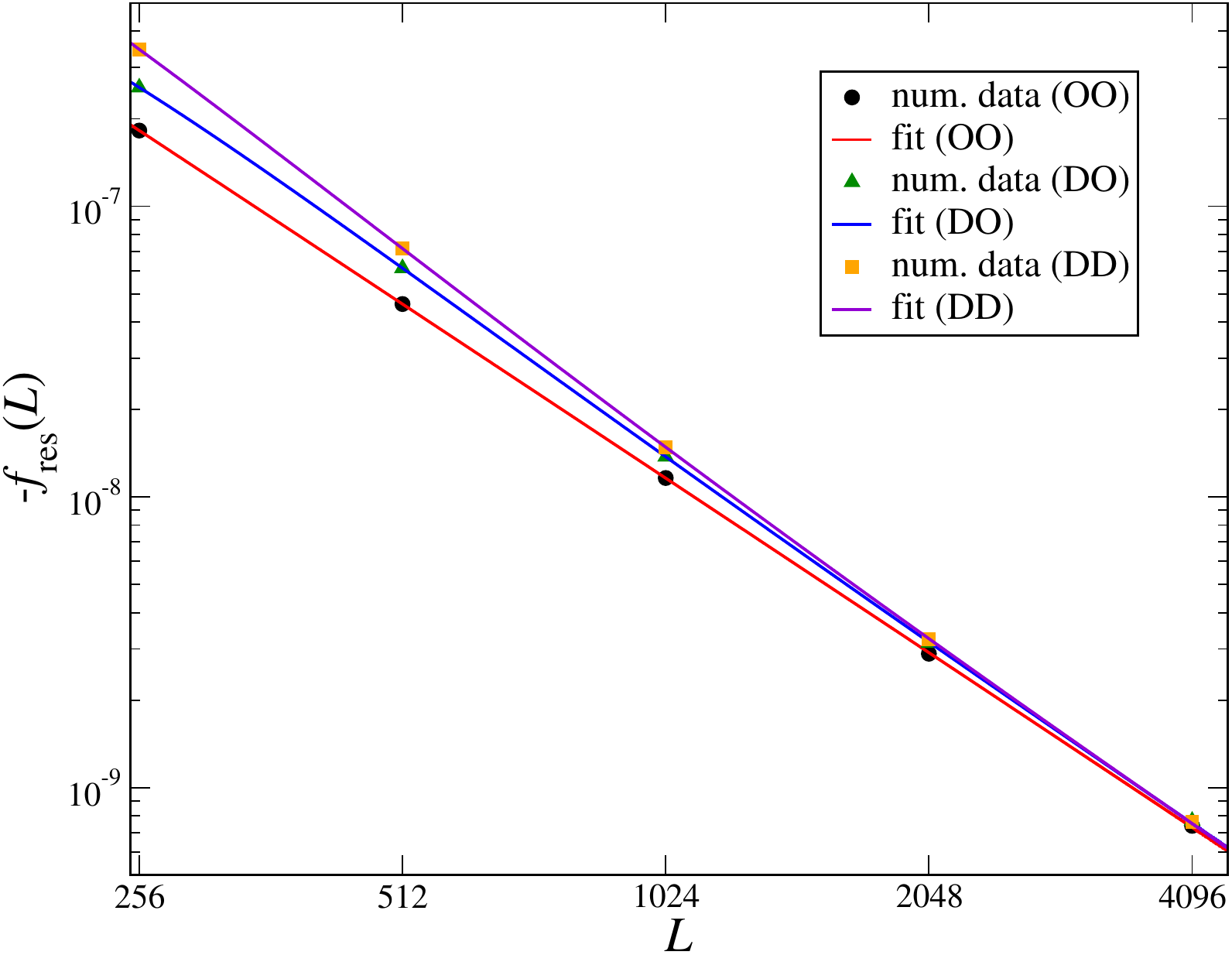}\hfill{}\includegraphics[clip,width=0.48\columnwidth]{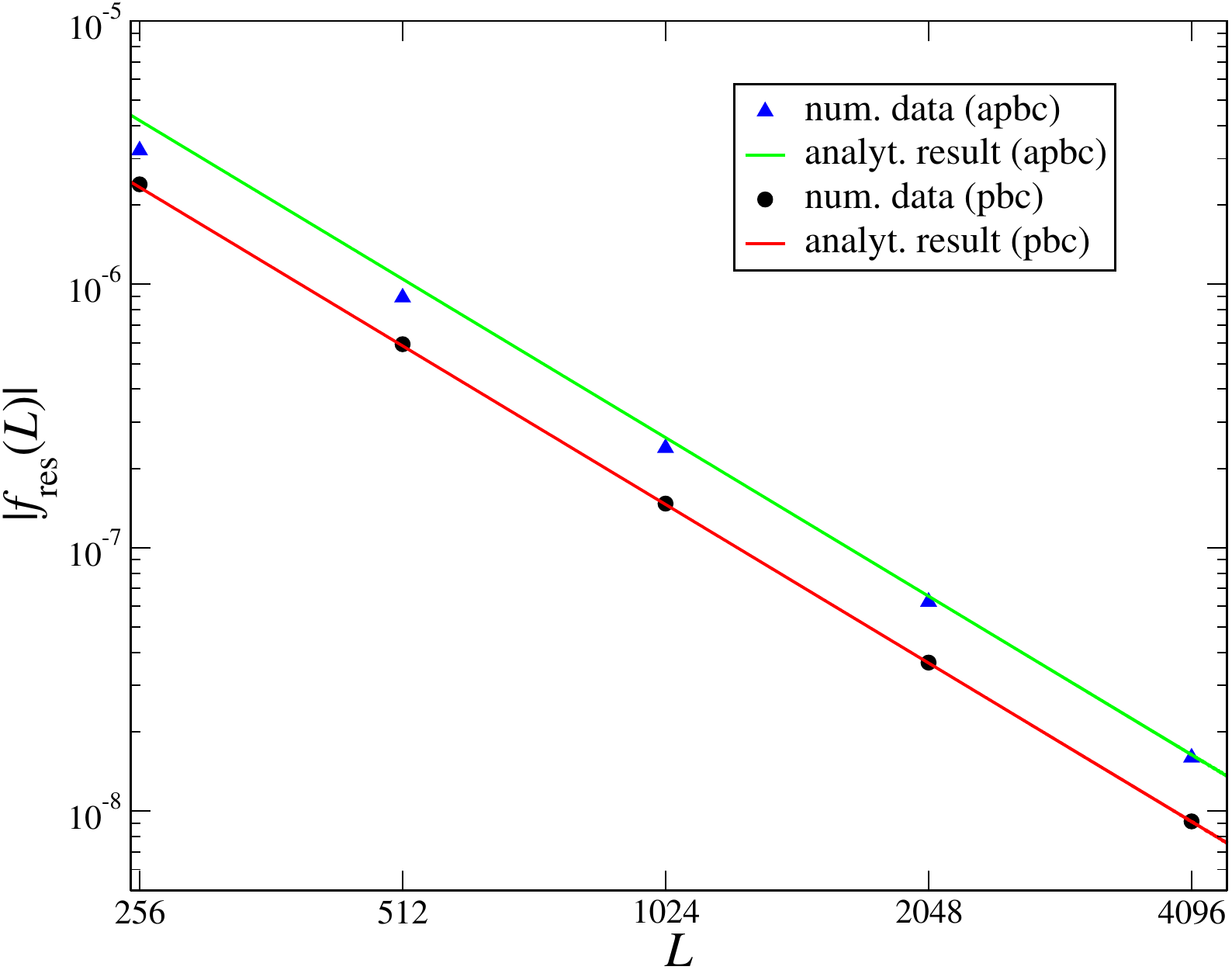}
\caption{Left: Numerical data with the corresponding fits for open-open (OO),
Dirichlet-open (DO) and Dirichlet-Dirichlet (DD) boundary conditions.
Right: Comparison between analytic and numeric results for periodic
(PBC) and antiperiodic (APBC) boundary conditions. \label{fig:fres}}
\end{figure}
To compute the Casimir amplitude from the evolution of the numerical
evaluated excess free energy with increasing system size we fit the
data using the \textit{ansatz} 
\begin{equation}
f_{\mathrm{ex}}=a_{0}+a_{2}L^{-2}+a_{3}L^{-3}+a_{4}L^{-4}+\ldots\label{eq:fit}
\end{equation}
where the factor $a_{2}$ will later be interpreted as the Casimir
amplitude. This fitting \textit{ansatz} is justified by the following:
From fundamental arguments one knows that the order $O(L^{-1})$ is
absent and the leading order is the order $O(L^{-2})$ \cite{FdG78,KD92b}.
Higher orders are caused by corrections to the leading scaling behavior
(for details see \cite{Weg72a}). Here we restrict ourselves to orders
up to $O(L^{-4})$. The constant term $a_{0}$ is caused by surface
contributions introduced by the considered boundary conditions and
is absent in the case of periodic and antiperiodic boundary conditions.
The numerical data and the associated fits for $\mathring{u}=1$ are
shown in Figure~\ref{fig:fres}(left) and the results for the fitting
parameters are assembled in Table~\ref{tab:FitTab}. 

Interpreting the fit parameter $a_{2}$ as the Casimir amplitude we
estimate the values 
\begin{eqnarray}
\Delta_{\mathrm{C,OO}} & = & -0.012(1)\\
\Delta_{\mathrm{C,DO}} & = & -0.012(3)\nonumber \\
\Delta_{\mathrm{C,DD}} & = & -0.012(3),\nonumber 
\end{eqnarray}
where the numbers in brackets give the error bars.

As anticipated the numerical values of the Casimir amplitudes lie
all together in a small range. They are equal up to the third decimal
place when we take the adjusted error bar into account. We see that
due to the fact that the boundary conditions are all in the same surface
universality class the evolutions of the excess free energies of the
three systems are asymptotically equivalent for large $L$. Looking
at the non-universal corrections to the leading scaling behavior namely
the terms of order $O(L^{-3})$ and $O(L^{-4})$ we see that their
weights are much higher in the DD and DO case compared to the OO case.

We checked the accuracy of our method by computing results for the
periodic and the antiperiodic case and comparing them to the analytical
results. Our fit parameters are also shown in Tab.~\ref{tab:FitTab}
and the numerical result are compared to the analytic results in Fig.~\ref{fig:fres}(right).
The analytical results for the Casimir amplitudes \cite{BDT00,Dan09}
are given by \begin{subequations} 
\begin{eqnarray}
\Delta_{\mathrm{C,PBC}} & = & -\frac{2\zeta(3)}{5\pi}\;\qquad\qquad=\;-0.153050\ldots\,,\\
\Delta_{\mathrm{C,APBC}} & = & \frac{\mathrm{Cl}_{2}(\pi/3)}{3}-\frac{\zeta(3)}{6\pi}\;=\;+0.274542\ldots\;.
\end{eqnarray}
\end{subequations}where $\mathrm{Cl}_{n}$ denotes the Clausen function.
For the case of periodic boundary conditions we find a perfect agreement
with the analytic results. For the case of antiperiodic boundary conditions
we find an asymptotic agreement for large $L$ because there are more
corrections due to the inhomogeneity introduced by this boundary condition.

\section{Summary and concluding remarks \label{sec:concl}}

We showed a new method for the computation of Casimir amplitudes in
the large-$n$ limit. This method can be applied to the case of slabs
with non translationally invariant boundary conditions as well as
for those with periodic and antiperiodic boundary conditions. We evaluated
the Casimir amplitudes for the cases of Dirichlet-Dirichlet, open-open
and Dirichlet-open boundary conditions and reproduced their asymptotical
equivalence in the limit of large systems. We find that the non-universal
corrections to the leading scaling behavior in the open-open case
are much smaller than in the other cases. From this we conclude that
evaluations of Casimir amplitudes following our method in this universality
class are simpler in this open-open case. We also confirmed the known
analytic result for the Casimir amplitudes in the case of periodic
and antiperiodic boundary conditions. 
\begin{acknowledgments}
We would like to thank H. W. Diehl for the inspiration and initial
ideas to work on this issue and for many constructive discussions.
We also want to thank Felix M. Schmidt and Hassan Chamati for discussions
and additional ideas.
\end{acknowledgments}
\bibliography{MyBank,bank,bank2,casi,Physik}

\C{
}
\end{document}